\documentclass[amsmath,amssymb,nofootinbib,showkeys]{revtex4}

\usepackage{graphicx}
\usepackage{dcolumn}
\usepackage{color}
\usepackage{bm}
\usepackage{amsmath,amssymb,graphicx}
\usepackage{epsfig}
\usepackage{enumerate}
\usepackage{array}
\usepackage{multirow}

\begin{document}
\title
{A reappraisal of Lagrangians with non-quadratic\\ velocity dependence and branched Hamiltonians}
\author{Bijan Bagchi\footnote{E-mail: bbagchi123@gmail.com}$^1$, Aritra Ghosh\footnote{E-mail: ag34@iitbbs.ac.in}$^2$, Miloslav Znojil\footnote{E-mail: znojil@ujf.cas.cz}\footnote{Corresponding author}$^{3,4,5}$}

\vspace{2mm}

\affiliation{$^{1}$Brainware University, Barasat, Kolkata, West Bengal 700125, India\\
$^{2}$School of Basic Sciences, Indian Institute of Technology Bhubaneswar, Jatni, Khurda, Odisha 752050, India\\
$^{3}$ The Czech Academy of Sciences, Nuclear Physics Institute, Hlavn\'i 130, 250 68 \v{R}e\v{z}, Czech Republic\\
$^{4}$
 Department of Physics, Faculty of
Science, University of Hradec Kr\'{a}lov\'{e}, Rokitansk\'{e}ho 62,
50003 Hradec Kr\'{a}lov\'{e},
 Czech Republic\\
 $^{5}$
Institute of System Science, Durban University of Technology,
Durban, South Africa
}

\vskip-2.8cm
\date{\today}
\vskip-0.9cm


\begin{abstract}
Time and again, non-conventional forms of Lagrangians with non-quadratic velocity dependence have found attention in the literature. For one thing, such Lagrangians have deep connections with several aspects of nonlinear dynamics including specifically the types of the Li\'{e}nard class; for another, very often the problem of their quantization opens up multiple branches of the corresponding Hamiltonians, ending up with the presence of singularities in the associated eigenfunctions. In this article, we furnish a brief review of the classical theory of such Lagrangians and the associated branched Hamiltonians, starting with the example of Li\'{e}nard-type systems. We then take up other cases where the Lagrangians depend upon the velocity with powers greater than two while still having a tractable mathematical structure, while also describing the associated branched Hamiltonians for such systems. For various examples, we emphasize upon the emergence of the notion of momentum-dependent mass in the theory of branched Hamiltonians.
 \end{abstract}


\maketitle

\section{Introduction}

During the past few decades, the study of non-conventional types of dynamical systems, in particular those which are controlled by Lagrangians that are not quadratic in the velocity has entered a new phase of intense development \cite{SW,SW1,SW2,MCZ}. Such Lagrangians lead to certain exotic Hamiltonians, commonly termed as \textit{branched Hamiltonians}, that have relevance in their applicability to problems of nonlinear dynamics pertaining to autonomous differential equations \cite{bspf, mitso}, and to certain exotic quantum-mechanical models especially in the context of non-hermitian parity-time ($\mathcal{PT}$)-symmetric schemes \cite{bender}, along with their relativistic counterparts \cite{bpm}. \\

A simple way to see how Lagrangians that are not quadratic in the velocity can lead to meaningful dynamical systems is to consider the following toy model \cite{benoy,carinena} (see also, Refs. \cite{CFN,CGR09}):
\begin{equation}\label{NS11}
    L(x,\dot{x}) = (\alpha x + \beta \dot{x})^{-1},
\end{equation} where \(\alpha\) and \(\beta\) are real numbers satisfying \(\alpha\beta > 0\). We may also require that \(\alpha x + \beta \dot{x} \neq 0\),
i.e., that the velocity phase space accessible to the system is 
defined as a subset of \(\mathbb{R}^2\).\\
 
Notice that the Lagrangian cannot be expressed as the difference between the kinetic and potential energies; such Lagrangians shall be referred to as \textit{nonstandard}, i.e., in this paper we will be adopting such a 
nomenclature in which the term `nonstandard Lagrangian'
would refer to a Lagrangian with a non-quadratic velocity dependence\footnote{A linear dependence on the velocity makes the Hessian matrix singular resulting in a singular Legendre transform while passing from the Lagrangian to the Hamiltonian formalism (see for example, Ref. \cite{dirac}). We do not address such cases here and deal with Lagrangians that have velocity dependence either in excess of quadratic powers or in inverse powers.}.\\

A direct computation reveals that the Euler-Lagrange equation is
\begin{equation}\label{lhodamped1}
  \ddot{x} + \gamma \dot{x} + \omega_0^2x=0,
\end{equation}
where \(\gamma=\frac{3 \alpha}{2 \beta}\) and \(\omega_0= \frac{\alpha}{\sqrt{2} \beta}\). (\ref{lhodamped1}) is just the harmonic oscillator in the presence of linear damping. We remind the reader that there is no time-independent Lagrangian of the `standard' kind from which one can reproduce (\ref{lhodamped1}) upon invoking the Euler-Lagrange equation\footnote{One could recover the damped oscillator from a standard Lagrangian by using a Rayleigh dissipation function \cite{goldsteincm}. Alternatively, one can consider the modified Euler-Lagrange equations from the Herglotz variational problem to describe the damped oscillator \cite{herglotz}. We do not consider such situations here.}. There exist various other families of nonstandard Lagrangians (giving rise to different dynamical systems) which look quite different from (\ref{NS11}); each family is endowed with their own intriguing features. However, the common theme is the existence of Lagrangians that are not quadratic in the velocity, thereby leading to a nonlinear relationship between the velocity and the momentum. It may be emphasized that a Lagrangian \(L = L(x,\dot{x})\) defined for a system whose configuration space is a subset of \(\mathbb{R}\) is called \textit{regular} if its Hessian with respect to the velocity is non-vanishing, i.e.,
\begin{equation}\label{regularcond}
\frac{\partial^2 L (x,\dot{x})}{\partial \dot{x}^2} \neq 0,
\end{equation} and is of constant sign allowing us to solve for the velocity \(\dot{x}\) in favor of the momentum \(p(x,\dot{x}) = \frac{\partial L(x,\dot{x})}{\partial \dot{x}}\), i.e., we can write \(\dot{x}(x,p)\). Thus, Lagrangians with a quadratic velocity dependence are regular and one can formulate a Hamiltonian description by the means of a Legendre transform. The condition (\ref{regularcond}) fails for Lagrangians that are linear in the velocity (see for instance, Ref. \cite{dirac}) but as mentioned earlier they won't be our concern here. Instead, we shall be looking at Lagrangians for which solving the equation \(p = (x,\dot{x})\) in favor of the \(\dot{x}\) leads to a non-unique solution, e.g., the appearance of a square root which gives rise to what will be called \textit{branching}. Such Lagrangians would not permit the construction of a Hamiltonian function in a unique way. \\

In the classical context, the problems associated with branched Hamiltonians and the ones that are inevitably posed after their quantization, were addressed by Shapere and Wilczek \cite{SW,SW1,SW2}. This has triggered off a series of papers by Curtright and Zachos \cite{CZ1,CZ2,CZ3,CZ4,CZ,C1} which were subsequently followed up by other works in a similar direction (see for example, Refs. \cite{bspf,bst,agp}). It bears mention that local branching is not so sufficient to ensure
integrability. In particular, finding an integrable differential equation having
solutions that are not locally finitely branched with a finitely-sheeted Riemann surface but
not yet identified through Painlev\'{e} analysis, is in itself an interesting open problem \cite{CZ1}.\\

Against this background, a new class of innovations
on the description and simulations of quantum dynamics emerged in relation to the specific role played by certain models constructed appropriately. Not quite unexpectedly, Hamiltonians which are multi-valued functions of momenta confront us with some typical insurmountable ambiguities of quantization. In such cases, the underlying
Lagrangian possesses time derivatives in excess of quadratic powers (or sometimes, inverse powers). The use of
these models leads, on both classical and quantum grounds, to the necessity of a re-evaluation
of the dynamical interpretation of the momentum, which, in principle, becomes a multi-valued
function of the velocity. It also needs to be pointed out that the
traditional approaches often do not always work as is the case with
certain $\mathcal{PT}$-symmetric complex potentials possessing real spectra \cite{BagG} or, on employing tractable non-local generalizations \cite{zno1}. \\

 In the context of nonlinear models, certain Li\'{e}nard-class systems present an intriguing feature
of the Hamiltonian in which the roles of the position and
momentum variables get exchanged with the emergence of the notion of a momentum-dependent mass \cite{agp,BGG,brp,ML,Bag2,ruby,modak}. Naturally, the presence of the
damping as is the case for Li\'{e}nard systems poses to be a problem whenever one tries to contemplate a quantization of the
model.
It is important to realize that the quantization is hard to tackle in the coordinate representation of the Schr\"{o}dinger equation but can be straightforwardly carried out in the momentum space \cite{BGG,ruby} (see also, Ref. \cite{Roos}). \\

Although much has been said about the quantum-mechanical formalisms, in this paper, we focus on the classical theory, (briefly) reviewing some aspects of nonstandard Lagrangians and the associated branched Hamiltonians. The theory is exemplified by focusing on various examples which include some systems of the Li\'{e}nard class, which are of great interest in the theory of dynamical systems. Apart from Li\'{e}nard systems, we discuss some interesting toy Lagrangians which contain time derivatives in excess of quadratic powers, leading to branched Hamiltonians. The basic features of the theory are discussed in the light of these examples. However, we begin with a discussion on some simple nonstandard Lagrangians which can be figured out via some guesswork in Section \ref{TLSec}. Following this, in Section \ref{Lsec} we discuss nonstandard Lagrangians and branched Hamiltonians in the context of Li\'{e}nard systems, wherein we outline a systematic derivation of the Lagrangians, provided the system admits a certain integrability condition. In Sections \ref{nslsec1} and \ref{nslsec2} we analyze various intriguing examples of Lagrangians in which time derivatives occur in excess of quadratic powers, while also discussing the associated Hamiltonians. We conclude with some remarks in Section \ref{dsec} and in Appendix A
where a few further aspects of the problem of quantization
are also discussed.

\section{Some illustrative examples}\label{TLSec}

\textbf{Example 1 --} Consider the following Lagrangian \cite{benoy,carinena}:
\begin{equation}\label{NSL12}
  L(x, \dot{x})=\frac{1}{ \alpha \mu(x) + \beta \dot{x}}, \quad\quad \alpha \dot{x} + \beta \mu(x) \neq 0,
\end{equation} where \(\mu(x)\) is a well-behaved function (typically a polynomial), while \(\alpha\) and \(\beta\) are real-valued and non-zero constant numbers. Obviously, it does not reveal the `standard' form as the difference between the kinetic and potential energies. However, the Euler-Lagrange equation gives \(\ddot{x} + f(x) \dot{x} + g(x) = 0\), with \(f(x)=\frac{3 \alpha \mu'(x)}{2 \beta}\) and \(g(x)= \frac{\alpha^2 \mu'(x)\mu(x)}{2 \beta^2}\), where for instance, picking \(\mu(x) = x\) gives the linearly-damped harmonic oscillator, while the choice \(\mu(x) = x^2\) implies \(f(x) \propto x\) and \(g(x) \propto x^3\). Lagrangians of this type [(\ref{NSL12})] are termed as \textit{reciprocal Lagrangians}. \\

\textbf{Example 2 --} Consider another form of Lagrangians classified by \cite{benoy}
\begin{equation}\label{LL}
L(x,\dot{x}) = \ln [\gamma \mu(x) + \delta \dot{x}], \quad \quad   \gamma \mu(x) + \delta \dot{x} > 0,
\end{equation} where \(\delta\) and \(\beta\) are real-valued and non-zero constant numbers. The Euler-Lagrange equation goes as \(\ddot{x} + f(x) \dot{x} + g(x) = 0\), with \(f(x) = \frac{2 \gamma \mu'(x)}{\delta}\) and \(g(x) = \frac{\gamma^2 \mu(x) \mu'(x)}{\delta^2}\). Lagrangians that look like (\ref{LL}) are termed as \textit{logarithmic Lagrangians}. The relation between logarithmic and reciprocal classes of Lagrangians has been explored in \cite{CFN} (see also, Ref. \cite{CGR09})\footnote{As with the system described by the Lagrangian (\ref{NS11}), the systems given by (\ref{NSL12}) and (\ref{LL}) are defined only on appropriate regions of \(\mathbb{R}^2\).}.\\

\textbf{Example 3 --} As another example, we point out that some equations that go as \(\ddot{x} + A(x,\dot{x}) \dot{x} + B(x,\dot{x}) = 0\), where \(A(x,\dot{x})\) and \(B(x,\dot{x})\) are suitable functions of \((x,\dot{x})\) can be derived from (reciprocal) Lagrangians that read
\begin{equation}\label{NSL0000}
  L(x,\dot{x})=\frac{1}{ \alpha \mu(x) + \beta \rho(\dot{x})},
\end{equation} such that \(\beta \rho''(\dot{x})[\alpha \mu(x) + \beta \rho(\dot{x})] \neq 2 \beta^2 \rho'(\dot{x})^2\). Specifically, the functions \(A(x,\dot{x})\) and \(B(x,\dot{x})\) are
\begin{equation}
    A(x,\dot{x}) = \frac{2 \alpha \beta \rho'(\dot{x}) \mu'(x) }{2 \beta^2 \rho'(\dot{x})^2 - \beta \rho''(\dot{x})[\alpha \mu(x) + \beta \rho(\dot{x})]},
    \end{equation}
    \begin{equation}
    B(x,\dot{x}) = \frac{\alpha \mu'(x) \big[ \alpha \mu(x) + \beta \rho(\dot{x}) \big]}{2 \beta^2 \rho'(\dot{x})^2 - \beta \rho''(\dot{x})[\alpha \mu(x) + \beta \rho(\dot{x})]}.
\end{equation}
However, there is a limited variety of differential equations that can be described by Lagrangians which may be guessed; in general, it is often not possible to systematically derive a Lagrangian from which a given differential equation may emerge as the Euler-Lagrange equation. In what follows, we describe Li\'{e}nard systems and demonstrate that if a certain integrability condition is satisfied, then one may systematically find nonstandard Lagrangians describing such systems.

\section{Li\'{e}nard systems}\label{Lsec}
A Li\'{e}nard system is a second-order ordinary differential equation that goes as
\begin{equation}\label{Lienard}
    \ddot{x} + f(x) \dot{x} + g(x) = 0,
\end{equation} where\footnote{Often it is sufficient to have \(f(x), g(x) \in C^2(U,\mathbb{R})\), where \(U \subset \mathbb{R}\).} \(f(x), g(x) \in C^\infty(\mathbb{R},\mathbb{R})\) can be suitably chosen. Interesting choices for \(f(x)\) and \(g(x)\) include \(f(x) = 1\) and \(g(x) = x\), which is just the damped linear oscillator, while the choice \(f(x) = (1- x^2)\) and \(g(x) = x\) gives the van der Pol oscillator \cite{VdP}, known to admit limit-cycle behavior due to the particular choice of \(f(x)\) \cite{str}. Another choice is \(f(x) = 1\) and \(g(x) = x^3\), for which we have the linearly-damped (nonlinear) Duffing oscillator (see for example, Refs. \cite{mic,Demina21}). It is noteworthy that in any case with \(f(x) \neq 0\), the system exhibits non-conservative dynamics because (\ref{Lienard}) does not stay invariant under the transformation \(t \rightarrow -t\), namely, time reversal. Further, oscillatory dynamics can be obtained if \(f(x)\) is an even function and if \(g(x)\) is odd; it follows from the fact that the overall force (the second and third terms of (\ref{Lienard})) should be odd under \(x \rightarrow -x\) in order to support oscillations \cite{mic}.

\subsection{Cheillini condition and nonstandard Lagrangians}\label{JLMSec}
Given a second-order differential equation,
the inverse problem of finding the Lagrangian has been the subject of much investigation \cite{whittaker,yan78,Leach1,Leach2,JLMLienard,mitra24} (see also, Ref. \cite{carinenaintegrability}). In particular, for Li\'{e}nard systems satisfying a certain integrability condition, one can find nonstandard Lagrangians from which they emerge as the Euler-Lagrange equation \cite{JLMLienard} (see also, Refs. \cite{agp,BGG}). The idea is to make use of the so-called Jacobi last multiplier which may be defined as in \cite{whittaker} (see Appendix (\ref{appA})). \\

In this manner, starting from an ordinary differential equation
\begin{equation}\label{ode}
    \ddot{x}=F(x,\dot{x}),
\end{equation} one defines the last multiplier $M$ as that which satisfies
\begin{equation}\label{Mdef1d}
    \frac{d\ln M}{dt}+\frac{\partial
F(x,\dot{x})}{\partial\dot{x}}=0.
\end{equation}
As has been discussed in Whittaker's classic textbook \cite{whittaker}, if a second-order differential equation such as (\ref{ode}) follows from the Euler-Lagrange equations, then the Lagrangian is related to the last multiplier as
\begin{equation}\label{abcd22}
    M=\frac{\partial^2L(x, \dot{x})}{\partial \dot{x}^2}.
\end{equation}
This allows one to determine the Lagrangian function for a given second-order differential equation, provided it admits a Lagrangian formalism.\\
 
For the Li\'{e}nard system, a formal solution for the last 
multiplier is found to be
\begin{equation}\label{abcd11}
    M(t,x)=\exp\left(\int f(x)
dt\right).
\end{equation} We may define a new nonlocal variable \(u\) as \cite{JLMLienard}
\begin{equation}\label{udefinition}
u = \dot{x} - \frac{g(x)}{\ell f(x)},
\end{equation} with \(\ell \neq -1\) which is determined from
\begin{equation}\label{cond1}
      \frac{d}{dx}\left(\frac{g(x)}{f(x)}\right)
  + \ell(\ell+1)f(x) = 0.
  \end{equation}
Then using Eqs. (\ref{udefinition}) and (\ref{cond1}) we have \( \dot{u}=\ell uf(x)\). In other words, if the condition (\ref{cond1}) is true the Li\'{e}nard system (\ref{Lienard}) can be expressed as the following system of first-order equations:
  \begin{equation}\label{lienardnewform}
      \dot{u}=\ell uf(x),\quad \quad \dot{x}=u+W(x),
  \end{equation} where $W(x)=\ell^{-1}g(x)/f(x)$. Since we have \(\dot{u}=\ell uf(x)\), using (\ref{abcd11}) we can write \(M = u^{1/\ell}\), which, upon using (\ref{abcd22}) gives us (up to a gauge function which can be ignored here) \(L \sim u^{\frac{2 \ell +1}{\ell}}\). In terms of \(x\) and \(\dot{x}\) the Lagrangian reads as \cite{agp}
  \begin{equation}\label{Lie1Lag}
  L(x,\dot{x}) = \frac{\ell^2}{(\ell+1)(2\ell+1)}\left(\dot{x}-\frac{1}{\ell}\frac{g(x)}{f(x)}\right)^{{\frac{2\ell+1}{\ell}}},
  \end{equation} which is of the nonstandard kind\footnote{It may be noted that one cannot at this stage set \(f(x) = 0\) in (\ref{udefinition}), (\ref{cond1}), (\ref{lienardnewform}), or (\ref{Lie1Lag}) to recover the conservative case. However, one can set \(f(x) = 0\) in (\ref{abcd11}) which gives \(M = 1\) and consequently, from (\ref{abcd22}) one has \(L(x,\dot{x}) = \frac{\dot{x}^2}{2} + J(x) \dot{x} + K(x)\), where \(K(x)\) is the (conservative) scalar potential while \(J(x)\) may be interpreted as a vector potential.}.\\

It is noteworthy that (\ref{cond1}) represents what is known as the Cheillini condition, allowing one to recast the Li\'{e}nard system in the form (\ref{lienardnewform}). Specifically, if \(f(x) = ax^\alpha\), then (\ref{cond1}) dictates that \(g(x)\) must satisfy the following differential equation (see also, Refs. \cite{CG19,CG24}):
  \begin{equation}\label{cond12}
      \frac{d}{dx}\left(x^{-\alpha}g(x)\right)
  + \ell(\ell+1)a^2 x^\alpha = 0.
  \end{equation}
A simple integration gives
\begin{equation}\label{gxchellinifunction}
g(x) = kx^{\alpha} - \frac{\ell (\ell + 1)a^2}{\alpha + 1} x^{2\alpha + 1},
\end{equation} where \(k \in \mathbb{R}\) is an integration constant. This gives the functional form of \(g(x)\) so as to satisfy the Cheillini condition. In the cases where the Cheillini condition is satisfied the Lagrangian is given by (\ref{Lie1Lag}).

\subsection{Hamiltonian aspects}
With the Lagrangian that describes the Li\'{e}nard system, we can move on to its Hamiltonian aspects. The conjugate momentum is found to be
\begin{equation}
p = \frac{\partial L}{\partial \dot{x}} = \frac{\ell}{\ell+1}\left(\dot{x}-\frac{1}{\ell}\frac{g(x)}{f(x)}\right)^{{(\ell+1)}/{\ell}}.
\end{equation}
Thus, the expression \(\dot{x} = \dot{x}(p)\) may be multi-valued, depending upon \(\ell\). It goes as
\begin{equation}\label{velocitylienardsolution}
\dot{x} = K(l) p^{\ell/(\ell+1)} +\frac{1}{\ell}\frac{g(x)}{f(x)},
\end{equation} where \(K(\ell)\) is some function of \(\ell\) and is a constant. Using the above form, the Hamiltonian is found to be
\begin{equation}\label{Hlienardgen}
H(x,p) = K(\ell) p^{\frac{2\ell + 1}{\ell + 1}} - \frac{g(x)}{\ell f(x)} p.
\end{equation}
Notice that if (\ref{velocitylienardsolution}) admits branching, then so does the Hamiltonian (\ref{Hlienardgen}). Below, we discuss a concrete example.

\subsection{A concrete example}
An illustrative example will help us understand the framework discussed above. In particular, this will also enable us to turn our attention to
the notion of momentum-dependent mass \cite{BGG,brp,ML}, a concept that has generated some interest in the recent times, especially within the quantum-mechanical physics-oriented model-building approaches.\\ 

We will consider the case where \(f(x) = x\) and \(g(x) = x - x^3\) \cite{agp}; (\ref{cond1}) is satisfied for \(\ell = 1, -2\). For the sake of clarity, we will consider each case separately.

\subsubsection{Case with \(\ell = 1\)}
\begin{figure}
\begin{center}
\includegraphics[scale=0.80]{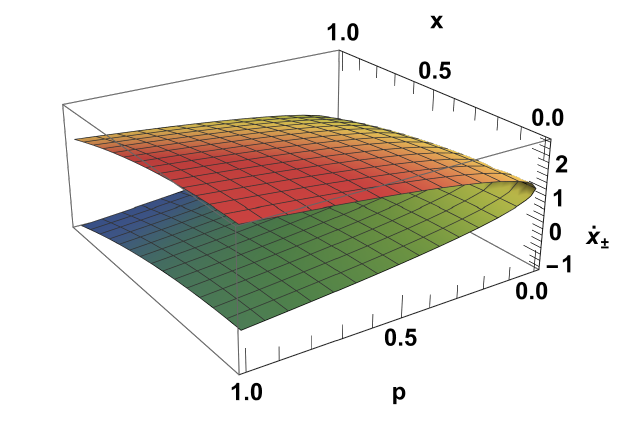}
\caption{Plot of \(\dot{x}_\pm = \dot{x}_\pm(x,p)\) for the case \(\ell = 1\). We have \(p \geq 0\) with the two branches meeting at \(p = 0\).}
\label{fig1}
\end{center}
\end{figure}

\begin{figure}
\begin{center}
\includegraphics[scale=0.80]{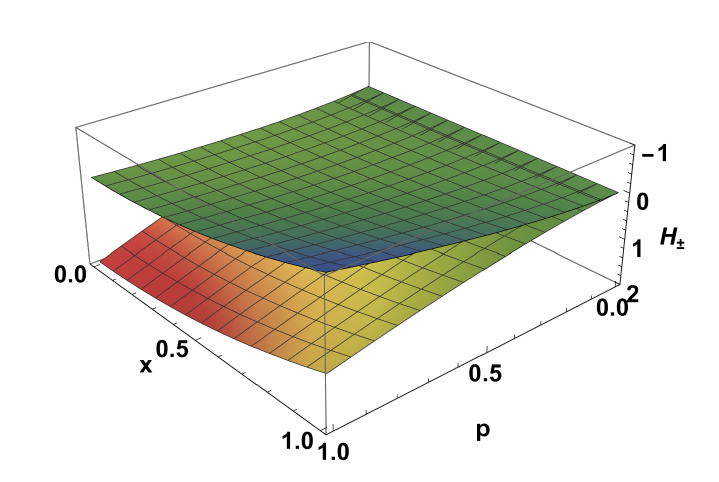}
\caption{Plot of the branched Hamiltonian \(H_\pm = H_\pm(x,p)\) arising for the case \(\ell = 1\), with \(p \geq 0\) and the two branches coalesce at \(p = 0\).}
\label{fig2}
\end{center}
\end{figure}

In the case of \(\,\ell = 1\) we have
\begin{equation}
p=p(x,\dot{x}) = \frac{1}{2} \left(\dot{x} + x^2 - 1 \right)^2 \quad \implies \quad \dot{x}= \dot{x}(x,p) = 1 - x^2 \pm \sqrt{2p}.
\end{equation} 
This points towards branching. Notice that branching originates from the nonlinear dependence between \(p\) and \(\dot{x}\) in the equation \(p = \frac{\partial L}{\partial \dot{x}}\). The corresponding branched Hamiltonians turn out to be
\begin{equation}
H_{\pm}(x,p) = p \left(1 - x^2 \pm \frac{2}{3} \sqrt{2p}\right) ,
\end{equation} exhibiting two distinct branches, where \( p \geq 0\). We have plotted the function \(\dot{x} = \dot{x} (p,x)\) in Fig. (\ref{fig1}), while Fig. (\ref{fig2}) shows a plot of the branched pair of Hamiltonians, \(H_{\pm}=H_{\pm}(x,p)\). The branches coalesce at \(p=0\).

\subsubsection{Case with \(\ell = -2\)}
In the case where \(\ell = - 2\), we have
\begin{equation}
p=p(x,\dot{x}) = 2  \left(\dot{x} + \frac{1 - x^2}{2} \right)^{1/2} \quad \implies \quad \dot{x} = \dot{x}(x,p) = \frac{p^2}{4} - \frac{(1-x^2)}{2},
\end{equation} implying that there is no branching, because \(\dot{x}\) can be extracted, uniquely, as a function of the momentum. A straightforward calculation reveals that the Hamiltonian turns out to be
\begin{equation}\label{HLienardexample2}
H(x,p) = \frac{p^3}{12} - \frac{p(1-x^2)}{2},
\end{equation}
wherein, there is only one branch.\\
 
\begin{figure}
\begin{center}
\includegraphics[scale=0.80]{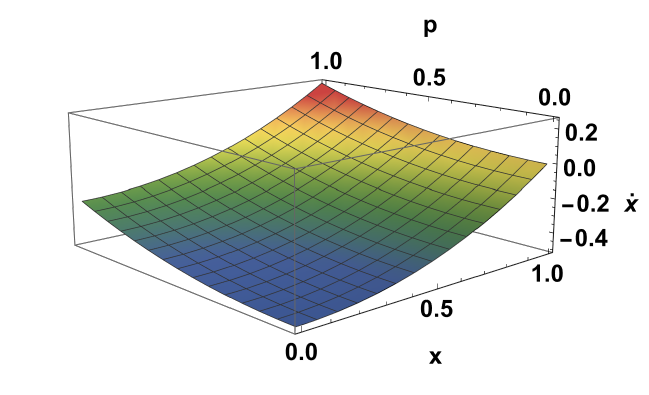}
\caption{Plot of \(\dot{x} = \dot{x}(x,p)\) for the case \(\ell = -2\). There are no branches.}
\label{fig3}
\end{center}
\end{figure}

\begin{figure}
\begin{center}
\includegraphics[scale=0.80]{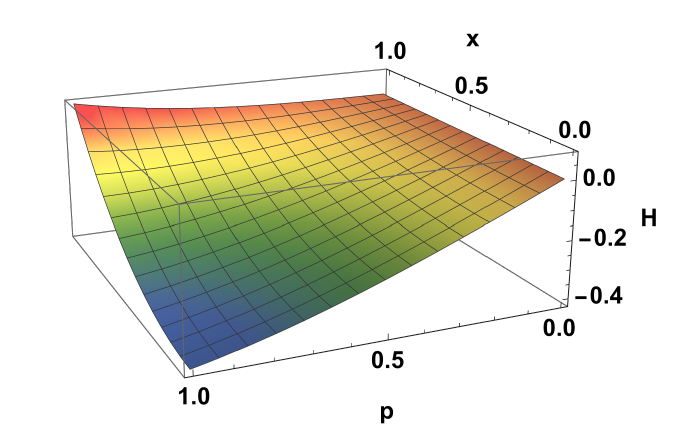}
\caption{Plot of the Hamiltonian \(H = H(x,p)\) arising for the case \(\ell = -2\), showing no branches.}
\label{fig4}
\end{center}
\end{figure}

In Figs. (\ref{fig3}) and (\ref{fig4}), we have plotted \(\dot{x} = \dot{x}(x,p)\) and \(H = H(x,p)\). An intriguing aspect of the Hamiltonian (\ref{HLienardexample2}) is that it may be expressed as
\begin{equation}
H(x,p) = \frac{x^2}{2 p^{-1}} + U(p), \quad\quad U(p) = \frac{p^3}{12} - \frac{p}{2}.
\label{uchos}
\end{equation}
This resembles a standard Hamiltonian, only with the roles of coordinate and momentum being interchanged. \\

It is then certainly tempting to interpret \(m(p) = p^{-1}\) as a \textit{momentum-dependent mass}. 
Also the quantization of such systems proceeds, in momentum space, often in the context referring to the notion of
momentum-dependent mass (see for example, Ref. \cite{BGG}). Still, in our particular model such a mass becomes singular (infinite) in the zero-momentum limit.
In the same limit, moreover,
also the potential itself is vanishing, i.e., a consistent
physical interpretation of
the system
would require a suitable regularization of the limiting process.\\

Marginally, let us add that one of the regularization recipes
which appeared applicable to 
the quantum version of
our very specific model (\ref{uchos})
has been proposed and tested in an older paper \cite{uchac}.
Based on an {\em ad hoc\,} complexification of the 
momentum
and on
a certain rather sophisticated
strong-coupling perturbation-expansion technique
the recipe has been even
found to provide the numerically fairly-reliable
spectra, in certain ranges of couplings at least.\\

\section {A generalized class of Lagrangians yielding branched Hamiltonians}\label{nslsec1}
 Let us note that if one is given a single-valued Lagrangian $L(x,v)$ and defines it according to
formula $L(x,v)=x^2-V(v)\,$
rather than according to the usual recipe
$L(x,v)=v^2-V(x)\,$ and, moreover,
if the $p$ or $v$ dependence is non-convex, then, as a result of employing the Legendre transformation, the
branched functions are always encountered
despite our having started from a single-valued Lagrangian or Hamiltonian function.

\subsection{The \(v^4\) model}\label{v4sec}
Shapere and Wilczek have discussed a concrete model depicting a non-convex nature of the Lagrangian which reads \cite{SW}
\begin{equation}\label{s1}
    L(v)=\frac{1}{4}v^4-\frac{\kappa}{2} v^2 \,,
\end{equation}
where \(v\) is the velocity\footnote{Now and in subsequent discussions, we will denote \(\dot{x} = v\).} and \(\kappa > 0\) is a coupling parameter. Corresponding to (\ref{s1}), the conjugate momentum is a cubic function in \(v\) that is given by
\begin{equation}\label{s2}
    p(v)=v^3-\kappa v\,.
\end{equation} Clearly, \(p\) is not monotonic in velocity, which may lead to branching. The corresponding Hamiltonian is obtained as
\begin{equation}\label{s3}
    H(p)=\frac{3}{4}v^{4}-\frac{\kappa}{2}v^2\,, \quad\quad v = v(p),
\end{equation}
which, like \(L(v)\), is also a multi-valued function (with cusps) in the conjugate momentum $p$, since each given $p$ corresponds to one or three values of $v$ as shown in (\ref{s2}).\\

For systems with a non-convex Lagrangian as sampled by (\ref{s1}), the routine construction of the corresponding Hamiltonian in conjugate momentum variable is not unique. An analogous incertitude is encountered in cosmology models \cite{ctv,nhm}, in generalized schemes of Einstein gravity which involve topological invariants, and in theories of higher-curvature gravity \cite{ctj}.\\

\subsection{Velocity-independent potentials}
Curtright and Zachos \cite{CZ} extended the analysis of \cite{SW} by considering a generalized class of non-quadratic Lagrangians that go as
\begin{equation}\label{one}
L(x,v) = C(v - 1)^{\frac{2k-1}{2k+1}}- V(x)\,, \quad  \quad C={\frac{2k+1}{2k-1}}\left(\frac{1}{4}\right)^{\frac{2}{2k+1}},
\end{equation}
where the traditional kinetic-energy term is replaced by a fractional function of the velocity variable $v$, and $V(x)$ represents a convenient local-interaction potential. The fractional powers facilitate the derivation of supersymmetric partner forms of the potential \'{a} \textit{la} Witten \cite{witt}. We remark that the $(2k+1)-$st
root of the first term in $L(x,v)$ is required to be real, and $>0$ or $<0$ for $v>1$ or $v<1$, respectively.\\

Let us focus on the case with $k=1$. Performing a Taylor expansion for $v$ near zero, we can write $L(x,v) \approx C(-1+\frac{v}{3}+\frac{v^2}{9}+O(v^3))-V(x)$. While the first term is merely a constant and the second term contributes to the boundary of the action and therefore does not influence the equations of motion, the third term yields the kinetic structure:

\begin{align}
 A =\int_{t_{1}}^{t_{2}}L(x,v)dt \approx  C\left(t_{2}-t_{1}+\frac{1}{3}(x(t_{2})-x(t_{1}))+\frac{1}{9}\int_{t_{1}}^{t_{2}}\,v^2dt+\int_{t_{1}}^{t_{2}}\,O(v^3)dt\right)-\int_{t_{1}}^{t_{2}}V(x)dt.
\end{align}
Thus, for small velocities, the action results in the usual Newtonian form of the equations of motion.\\

For the large velocities, on the other hand,
we have a less trivial scenario which leads (for finite, positive-integer values of $k$) to a non-convex function of $v$. The curvature term corresponding to the quantity $\frac{\partial^2L}{\partial v^2}$ changes sign at the point $v=1$. Thus, $L(x,v)$ may be interpreted as a single pair of convex functions that have been judiciously pieced together. Now, from the Lagrangian $(\ref{one})$, the canonical momentum can be 
calculated, 
\begin{equation}
    p=p(v) = \left(\frac{1}{4}\right)^\frac{2}{2k+1}\frac{1}{(v-1)^\frac{2}{2k+1}}\,.
\end{equation}
Inverting the relation we observe that the velocity variable $v(p)$ emerges as a double-valued function of $p$:
\begin{equation}
   v= v_{\pm}(p)=1{\mp}\frac{1}{4}\left(\frac{1}{\sqrt p}\right)^{(2k+1)}.
\end{equation}
Corresponding to the two signs above, a pair branches of the Hamiltonian, namely, $H_{\pm}(x,p)$ will appear. Specifically, for any positive-integer value of $k$, these may be identified to be
\begin{equation}\label{Hzachoes1}
    H_{\pm}(x,p)=p{\pm}\frac{1}{4k-2}\left(\frac{1}{\sqrt p}\right)^{2k-1} +V(x)\,.
\end{equation}
From a classical perspective, in order to avoid an imaginary $v(p)$, one needs to address a nonnegative $p$. This in turn implies that the slope $\frac{\partial L}{\partial v}$ is always positive. It is interesting to note that for the $k=1$ case we are led to the quantum-mechanical supersymmetric structure for the difference $ H_\pm(x,p)-V(x)$, which reads $p\pm\frac{1}{2\sqrt{p}}$, in the momentum space. The associated spectral properties have been analyzed in the literature \cite{ruby, BGG}. \\

We end our discussion on this example by noting that in the special case where \(V(x) = x^2\), the branched Hamiltonian is \(H_{\pm}(x,p) = x^2 + U_\pm (p)\), wherein it appears as if the roles of the coordinate and the momentum have been interchanged, with \(U_\pm(p)\) being a momentum-dependent potential that exhibits two branches.

\subsection{Velocity-dependent potentials}

Lines of force can be ascertained with the help of velocity-dependent potentials which ensure that particles take to certain specified paths \cite{coff,goldsteincm}. In electrodynamics, the field vectors $\vec{E}$ and $\vec{B}$ can be determined given such a potential function when the trajectories of a charged particle's motion are specified. In the present context,  we  proceed to set up an extended scheme where the Lagrangian depends upon a velocity-dependent potential $\mathcal{V}(x,v)$ in the manner as given by \cite{bst,Bag2}
\begin{equation}\label{vdp1}
L(x,v) = C(v - 1)^{\frac{2k-1}{2k+1}}- \mathcal{V}(x,v)\,, \quad C={\frac{2k+1}{2k-1}}\left(\frac{1}{4}\right)^{\frac{2}{2k+1}},
\end{equation}\\
where $\mathcal{V}(x,v)$ is assumed to be given in a separable form, i.e., $\mathcal{V}(x,v)=U(v)+V(x)$; here $U(v)$ and $V(x)$ are well-behaved functions of $v$ and $x$, respectively. Using the standard definition of the canonical momentum, we find its form to be
\begin{equation}\label{pvexample2}
  p=  p(v)=\left(\frac{1}{4}\right)^{\frac{2}{2k+1}} (v-1)^{-\frac{2}{2k+1}}-U'(v)\,.
\end{equation}
The complexity of the right side does not facilitate an easy inversion of the above relation that would reveal the multi-valued nature of velocity in a closed, tractable form. Nevertheless, the associated branches of the Hamiltonian can be straightforwardly written down on employing the Legendre transform as
\begin{equation}
     H_\pm(x,p)=p\pm\frac{1}{4}[p+U'(v)]^{-\frac{2k-1}{2}}\left(\frac{2k+1}{2k-1}-p[p+U'(v)]^{-1}\right)+\mathcal{V}(x,v)\,, \quad\quad v = v(p).
\end{equation}
Unfortunately, since a Hamiltonian has to be a function of the coordinate and its corresponding canonical momentum, the generality of the form of $H_\pm(x,p)$ as derived above is of little use unless we have an explicit inversion of (\ref{pvexample2}) giving \(v = v(p)\). We therefore have to go for the specific cases of $k$ and \(U(v)\).

\subsubsection{A special case}
 Indeed, the case $k=1$  proves to be particularly worthwhile to understand the spectral properties of the Hamiltonian. It corresponds to the Lagrangian as given by

\begin{equation}
    L(x,v)=3\left(\frac{1}{4}\right)^{\frac{2}{3}}(v-1)^{\frac{1}{3}}-U(v)-V(x)\,.
\end{equation}

A sample choice for $U(v)$ could be \cite{bst}
\begin{equation}
    U(v)=\lambda v+3\delta (v-1)^{\frac{1}{3}}\,,
\end{equation}\\
in which $\lambda(\geq 0)$ and $\delta( < 4^{-\frac{2}{3}})$ are suitable real constants. The presence of the parameter $\delta$ scales the kinetic-energy term in the Lagrangian.
The canonical momentum $p$ is now given by
\begin{equation}
p=p(v)=\mu (v-1)^{-\frac{2}{3}}-\lambda\,,
 \end{equation}
  where the quantity $\mu=4^{-\frac{2}{3}}-\delta>0$. We are therefore led to a pair of relations for the velocity depending on $p$:

 \begin{equation}
   v=  v_\pm(p)=1\mp\mu ^{\frac{3}{2}}(p+\lambda)^{-\frac{3}{2}}\,.
 \end{equation}\\
  In consequence, we find two branches of the Hamiltonian which are expressible as

  \begin{equation}
      H_{\pm}(x,p)  = (p + \lambda) \pm \frac{2\gamma}{\sqrt{p + \lambda}} + V(x),
  \end{equation}
 where $\mu^{3/2}$ has been replaced by $\gamma$. As a final comment, the special case corresponding to $\lambda=0$ and $\gamma=1/4$ conforms to the Hamiltonian (\ref{Hzachoes1}) advanced in \cite{CZ}.

\section{Three more forms of Hamiltonians}\label{nslsec2}

\subsection{Higher-power
Lagrangians}

As an extension of (\ref{one}), the following higher-power
Lagrangian was proposed in \cite{bspf}:
\begin{equation}\label{nine}
L(x,v) = C(v + \sigma(x))^{\frac{2m+1}{2m-1}}-\delta,\quad \Lambda =
 \left(\frac{1-2m}{1+2m}\right)(\delta)^{\frac{2}{1-2m}},
 \quad \delta> 0,
\end{equation}
where we notice that the coefficient $\Lambda$ is non-negative for $0 \leq m < \frac{1}{2}$. The main difference from (\ref{one}) is in the choice of a general
function $\sigma(x)$ in place of $\sigma(x)=-1$ as in (\ref{one}). The other point is that the inverse exponent with respect to the model of Curtright and Zachos \cite{CZ} has been taken for convenience of calculus. We have omitted the explicit potential function assuming that the interaction re-appears in a more natural manner
via a suitable choice of an auxiliary free parameter $\delta$ and that of a nontrivial function $\sigma(x)$.  As long as our Lagrangian $L(x,v)$ is of
a nonstandard type, we will not feel disturbed by the absence of the
explicit potential  $V(x)$.\\

For this particular model, the canonical momentum reads as
 \begin{equation}\label{ten}
 p = p(x,v) = -(\delta)^{\frac{2}{1-2m}}(v+\sigma(x))^{\frac{2}{2m-1}},
 \end{equation}
and a simple inversion yields
 \begin{equation}\label{eleven}
 v = v_\pm(x,p)=-\sigma(x)+\delta\big(\pm\sqrt{-p}\big)^{2m-1}\,.
 \end{equation}
This means, the Hamiltonian is obtained to be
 \begin{equation}\label{twelve}
 H_{\pm}(x,p)=(-p) \sigma(x)-
\frac{2\delta}{{2m+1}}\big(\pm\sqrt{-p}\big)^{2m+1}  + \delta\,.
 \end{equation}

 \subsubsection{Special case}
 The specific case with $m=0$ is of interest as it allows us to easily derive the (double-valued) velocity profile which reads as
\begin{equation}\label{thirteen}
v = v_{\pm}(x,p)= -\sigma(x) \pm \frac{\delta}{\sqrt{-p}}\,,
\end{equation}
implying that the Hamiltonian branches out into components:
\begin{equation}\label{fourteen}
H_{\pm}(x,p)= (-p)\sigma(x)\mp{2\delta}\sqrt{-p}+\delta\,.
\end{equation}
The nature of the two Hamiltonians depends on the sign of $p$. Once we specify the following choice of $\sigma(x)$, namely,
 \begin{equation}\label{fifteen}
 \sigma(x)=\frac{\lambda}{2}x^{2}
 +\frac{9\lambda^{2}}{2k^{2}},  \quad \lambda > 0,
 \end{equation} together with the choice \(\delta = \frac{9\lambda^2}{2k^2}\), then upon imposing a simple translation $p\rightarrow \frac{2k}{3\lambda}p- 1$, the Hamiltonians $H_{\pm}$
acquire the forms that go as
 \begin{equation}\label{sixteen}
 H_{\pm}(x,p)=\frac{9\lambda^{2}}{2k^{2}}
 \left[2  \mp 2\left(1-\frac{2kp}{3\lambda}
 \right)^{\frac{1}{2}}
 +\frac{k^{2}x^{2}}{9\lambda}
 -\frac{2kp}{3\lambda}-\frac{2k^{3}x^{2}p}{27\lambda^{2}}\right]\,.
 \end{equation}
These are readily identifiable as a set of plausible Hamiltonians representing a nonlinear Li\'{e}nard
system \cite{agp,BGG,ML}. The appearance of the coordinate-momentum coupling is noteworthy, and leads us to the notion of a momentum-dependent mass as
\begin{equation}
H_\pm (x,p) = \frac{x^2}{2 m(p)} + U_\pm(p), \quad\quad m(p) = \left[\lambda - \frac{2kp}{3}\right]^{-1}, \quad\quad U_\pm (p) = \frac{9\lambda^{2}}{2k^{2}}
 \left[2  \mp 2\left(1-\frac{2kp}{3\lambda}
 \right)^{\frac{1}{2}} -\frac{2kp}{3\lambda}\right].
\end{equation}
From a classical perspective, the momentum $p$ is needed to be restricted to the range
$-\infty<p\leq\frac{3\lambda}{2k}$  to account for the
physical properties of the system in the real space; this also ensures that the momentum-dependent mass is positive and finite. However, because of a branch-point singularity at $p=\frac{3\lambda}{2k}$, a thorough analytical study of $H_{\pm}(x,p)$  becomes greatly involved. Observe that when
$p=\frac{3\lambda}{2k}$, we find the coincidence of the two Hamiltonians $H_{\pm}(x,p)$.

\subsection{Rational-function
Lagrangians}

In another characteristic example, let us pick up an illustration where $L(x,v)$ is of the reciprocal kind and is defined to be \cite{bspf}
\begin{equation}\label{seventeen}
L(x,v)=\frac{1}{s}\left(\frac{1}{3}sx^{2}+\frac{3}{s}\lambda - v\right)^{-1},
\end{equation}
where $s$ is a real parameter. The canonical momentum comes out as
 \begin{equation}\label{eighteen}
p =  p(x,v)=
 \frac{1}{s}\left(\frac{1}{3}sx^{2}+\frac{3}{s}\lambda -v\right)^{-2}\,,
 \end{equation}
which, when inverted yields
 \begin{equation}\label{nineteen}
v= v_\pm(x,p)= \frac{1}{3}sx^{2}+\frac{3}{s}\lambda \pm \frac{1}{\sqrt{sp}}\,.
 \end{equation}
The accompanying
Hamiltonian corresponding to the above Lagrangian has two branches:
 \begin{equation}\label{twenty}
 H_{\pm}(x,p)=\frac{s}{3}x^{2}p
 +\frac{3}{s}\lambda p\pm2\sqrt{\frac{p}{s}}\,.
 \end{equation}
It should be remarked that as \(\lambda \rightarrow 0\), (\ref{seventeen}) is just the trial Lagrangian (\ref{NSL0000}), for the choice \(\mu(x) = ax^2\) and under a suitable identification of the constant parameters. We end by noting that (\ref{twenty}) can be re-expressed as a model,
\begin{equation}
H_\pm(x,p) = \frac{x^2}{2m(p)} + U_\pm (p), \quad \quad m(p) = \frac{3}{2 sp}, \quad \quad U_\pm(p) = \frac{3}{s}\lambda p\pm2\sqrt{\frac{p}{s}},
\end{equation}
with momentum-dependent mass.\\

\subsection{Relativistic free particle}\label{relsec}
As a final example of dynamics due to non-quadratic Lagrangians, let us re-examine the much-studied problem of a relativistic (free) particle which is described by the Lagrangian
\begin{equation}
L(v) = - mc \sqrt{c^2 - v^2}, \quad \quad v < c.
\end{equation}
The conjugate momentum is obtained as
\begin{equation}
p = p(v) = \frac{\partial L(v)}{\partial v} = \frac{mcv}{\sqrt{c^2 - v^2}}.
\end{equation}This implies that \(p^2c^2 - p^2 v^2 - m^2 c^2 v^2 = 0\). Solving for the velocity gives
\begin{equation}
v = v_\pm(p) = \pm \frac{pc}{\sqrt{p^2 + m^2 c^2}}.
\end{equation}
Thus, the Hamiltonian reads
\begin{equation}
H_\pm (p) = \frac{c(\pm p^2 - m^2 c^2)}{\sqrt{p^2 + m^2 c^2}}.
\end{equation}
In particular, one may consider the ultrarelativistic limit (\(v \approx c\)) in which \(H_\pm (p) = \pm pc \). Related to the above example, the reader is referred to Ref. \cite{example1} for the example of a spinning particle with the Lagrangian being non-quadratic both in the position and spin variables. Another physically-interesting example is that of an axially-symmetric charged body in an electromagnetic field which is governed by Euler-Poisson equations \cite{example2}.

\section{Concluding remarks}\label{dsec}

In the present treatment on the existence of nonstandard Lagrangians, we emphasized, first of all, the existence of certain
unusual aspects of their relationship with the associated branched Hamiltonians. Various different examples were discussed; in all of them, the velocity dependence of Lagrangian was not of (homogenous) degree two but contained either powers larger than two or negative powers. This resulted in a nonlinear relationship between the generalized velocity and the conjugate momentum, leading to a multi-valued behavior of the velocity when solved as a function of the momentum (and perhaps the coordinate). \\

We observed that in the description of Hamiltonians emerging from nonstandard Lagrangians, the notion of momentum-dependent mass is often encountered. It is then as if the coordinate of the particle played the role of momentum and vice versa, with a function of the momentum variable appearing as an `effective mass' describing the system.
Such systems can be quantized straightforwardly in the momentum space \cite{BGG,Roos,ruby}.
Naturally, this reopens a few mathematically-deeper
questions concerning their quantization. Indeed, the technicalities of canonical quantization
can be perceived as widely assessed in the literature
(see for example, Ref. \cite{kla}) wherein it is not infrequent
to encounter certain fundamental difficulties.
For example, in certain `anomalous' quantum
systems with non-hermitian Hamiltonians
supporting real eigenvalues, it has been shown that the
quantum wave functions themselves could still, in finite time, diverge \cite{eh}.
Moreover, after one admits the unusual forms of the Hamiltonians
characterized, typically, by the popular
parity-time symmetry ($\mathcal{PT}$-symmetry
-- see for example, Refs. \cite{rn, fr} for a
pedagogic and introductory discussion
on such specific variants of non-self-adjoint models),
the anomalies may occur even when the
$\mathcal{PT}$-symmetry itself remains unbroken.\\

Several unusual forms of the latter anomalies
may appear in
both the spectra and eigenfunctions,
materialized as the Kato's exceptional points \cite{kat, heiss}
or the so-called spectral singularities \cite{ply}. In particular, exceptional points can be regarded as a typical feature of non-hermitian systems related to a branch-point singularity where two or more discrete eigenvalues, real or complex,
and corresponding to two different quantum states, along with their accompanying eigenfunctions, coalesce \cite{zno, ff, BGS}.\\

Naturally, the possible relevance of the latter anomalies
in the quantum systems controlled by the branched Hamiltonians
is more than obvious. One only has to emphasize the difference
between the systems characterized by the unitary and non-unitary evolution.
In the former case, indeed, one is mainly interested in
the description of the systems of stable bound states. In the latter setting,
the scope of the theory is broader; the states are
resonant and unstable in general. In the related models, one
deals with Hamiltonians that are manifestly
non-hermitian and which undergo non-unitary quantum evolution; they generally represent open systems with balanced gain and loss \cite{moi, rot1}. Exceptional points occur there
as experimentally-measurable phenomena.
In this connection, it is also relevant to point out the
occurrence of certain theoretical anomalies like the possible
breakdown
of the adiabatic theorem \cite{kk}, or the feature of stability-loss delay \cite{tj},
etc. In all of these contexts, one encounters the possibility of interpreting branched Hamiltonians
as an innovative theoretical tool admitting
a coalescence of the branched pairs of operators at an exceptional point. Thus,
preliminarily, let us conclude that
the (related) possible
innovative paths towards quantization look truly promising.

\section*{Acknowledgments}
We thank Prof. Anindya Ghose Choudhury for discussions and for his interest in this work. B.B. thanks Brainware University for infrastructural support. A.G. thanks the Ministry of Education (MoE), Government of India, for financial support in the form of a Prime Minister’s Research Fellowship (ID: 1200454). M.Z. is financially supported by the Faculty of Science of UHK.

\appendix

\section{Jacobi last multiplier}\label{appA}
The Jacobi last multiplier \cite{whittaker} is a very useful tool in the mathematics of dynamical systems. On one hand, given a dynamical system on an \(m\)-dimensional phase space with \((m-2)\) linearly-independent and known first integrals, it allows one to determine the `last', i.e., \((m-1)\)th independent first integral; on the other hand, given a second-order (ordinary) differential equation, it facilitates the computation of a suitable Lagrangian function which describes the dynamics via the Euler-Lagrange equation \cite{whittaker,yan78,Leach1,Leach2}. Thus, the last multiplier is intimately related with the integrability properties of a dynamical system. Naively, one can define it as follows. Given a vector field \(X\) on the phase space, the Jacobi last multiplier \(M\) is a factor such that \(MX\) has zero divergence. For Hamiltonian systems for which Liouville's theorem holds, i.e., \({\rm div}\cdot X = 0\), the last multiplier is just a constant number as \(X\) is already divergence free. On the other hand, \(M\) assumes a more non-trivial form if \(X\) has non-vanishing divergence, i.e., if \({\rm div}\cdot X \neq 0\), say, for the Li\'enard equation, which is dissipative and hence, is not volume conserving (see for example, Refs. \cite{CG19,CG24}). \\

Consider a vector field \(X\); in local coordinates \(\{x_j\}\), where \(j = 1,2,\cdots,m\), one can write its components as \(\dot{x}_j = X_j(\{x_j\})\) which define the dynamical system as a system of first order equations. Further, let there be a certain number of first integrals \((F_1, F_2,\cdots, F_k)\), where \(k < m\). For any open subset \(\Omega \subset \mathbb{R}^m\), we define the Jacobi last multiplier to be a function \(M: \mathbb{R}^m \rightarrow \mathbb{R}\) which is non-negative and defines an invariant measure \(\int_\Omega M d^m x\), i.e.,
\begin{equation}
\int_\Omega M d x_1 \wedge  d x_2 \wedge \cdots \wedge d x_m = \int_{\phi_t({\Omega})} M \frac{\partial(x_1,x_2,\cdots,x_k,x_{k+1},\cdots, x_m)}{\partial(F_1,F_2,\cdots,F_k, x_{k+1},\cdots, x_m)} d F_1 \wedge dF_2 \wedge  \cdots \wedge d F_k \wedge dx_{k+1} \wedge \cdots dx_m,
\end{equation} where \(\phi_t(\Omega)\) is the transformation of the region \(\Omega\) under the flow of \(X\). Notice that it is necessary that \((F_1,F_2,\cdots,F_k)\) must be independent, ensuring that \(dF_1 \wedge dF_2 \wedge \cdots \wedge dF_k \neq 0\). The above-mentioned invariance condition leads to the equation
\begin{equation}
\frac{d}{dt} \Bigg( M \frac{\partial(x_1,x_2,\cdots,x_k,x_{k+1},\cdots, x_m)}{\partial(F_1,F_2,\cdots,F_k, x_{k+1},\cdots, x_m)}  \Bigg) = 0,
\end{equation} which, upon employing the chain rule gives
\begin{equation}
\frac{dM}{dt}  \frac{\partial(x_1,x_2,\cdots,x_k,x_{k+1},\cdots, x_m)}{\partial(F_1,F_2,\cdots,F_k, x_{k+1},\cdots, x_m)}  + M \sum_{j=1}^m \frac{\partial X_j}{\partial x_j}  \frac{\partial(x_1,x_2,\cdots,x_k,x_{k+1},\cdots, x_m)}{\partial(F_1,F_2,\cdots,F_k, x_{k+1},\cdots, x_m)}= 0.
\end{equation} This is just equivalent to
\begin{equation}\label{meqn}
\frac{d}{dt} \ln M + \sum_{j=1}^m \frac{\partial X_j}{\partial x_j} = 0,
\end{equation} which coincides with (\ref{Mdef1d}) as appropriate for the system (\ref{ode}). Notice that \(\sum_{j=1}^m \frac{\partial X_j}{\partial x_j}\) is just the divergence of \(X\) and therefore, \(M = 1\) or some constant if the vector field has zero divergence. \\

Now having defined the Jacobi last multiplier, let us demonstrate its use in deriving Lagrangians for a second-order ordinary differential equation with a two-dimensional phase space. Consider the system (\ref{ode}). If it is derivable from the Euler-Lagrange equation
\begin{equation}
\frac{d}{dt}\bigg( \frac{\partial L(x,\dot{x})}{\partial \dot{x}}\bigg) = \frac{\partial L(x,\dot{x})}{\partial x},
\end{equation}
or,
\begin{equation}
\frac{\partial^2 L(x,\dot{x})}{\partial \dot{x}^2} \ddot{x} + \frac{\partial^2 L(x,\dot{x})}{\partial \dot{x} \partial x} \dot{x} -  \frac{\partial L(x,\dot{x})}{\partial x} = 0,
\end{equation} then, from (\ref{ode}) one should have
\begin{equation}
\frac{\partial^2 L(x,\dot{x})}{\partial \dot{x}^2} F(x,\dot{x}) + \frac{\partial^2 L(x,\dot{x})}{\partial \dot{x} \partial x} \dot{x} -  \frac{\partial L(x,\dot{x})}{\partial x} = 0.
\end{equation}
Differentiating both sides with respect to \(\dot{x}\) gives
\begin{equation}\label{jlmprecondition}
\frac{\partial}{\partial \dot{x}} \bigg(\frac{\partial^2 L(x,\dot{x})}{\partial \dot{x}^2} F(x,\dot{x}) \bigg) + \frac{\partial^3 L(x,\dot{x})}{\partial \dot{x}^2 \partial x} \dot{x} = 0.
\end{equation}
Defining \(\Sigma(x,\dot{x}) = \frac{\partial^2 L(x,\dot{x})}{\partial \dot{x}^2}\), (\ref{jlmprecondition}) gives
\begin{equation}
\frac{\partial}{\partial \dot{x}} (\Sigma(x,\dot{x}) F(x,\dot{x})) + \frac{\partial}{\partial x} (\Sigma(x,\dot{x}) \dot{x}) = 0.
\end{equation} But this is just the equation (\ref{meqn}) of the last multiplier (if \(M\) has no explicit time dependence), i.e., we should identify \(\Sigma(x,\dot{x}) = M(x,\dot{x})\) and this readily gives (\ref{abcd22}). See Refs. \cite{whittaker,yan78,Leach1,Leach2} for more details.

 \end{document}